\documentclass[final,3p,times,twocolumn]{elsarticle}
 
\usepackage{lineno}
\modulolinenumbers[5]

\usepackage[colorlinks=true]{hyperref}

\journal{Advances in Space Research}

\biboptions{authoryear}

\newcommand\apjl{{Astrophys.\ J.}}%
\newcommand\aap{{Astron. Astrophys.}}%
\newcommand\nat{{Nature}}%
\newcommand\pasj{{PASJ}}%
\newcommand\procspie{{Proc.~SPIE}}%

\begin{document}

\begin{frontmatter}

\title{In-orbit Operation and Performance of the CubeSat Soft X-ray Polarimeter \textit{PolarLight}}

\author[thu-dep]{Hong Li}
\author[thu-dep]{Xiangyun Long}
\author[thu-doa,thu-dep]{Hua Feng\corref{cor}}
\cortext[cor]{Corresponding author}
\ead{hfeng@tsinghua.edu.cn}
\author[thu-dep]{Qiong Wu}
\author[thu-dep]{Jiahui Huang}
\author[ihep]{Weichun Jiang}
\author[infn-pisa]{Massimo Minuti}
\author[thu-dep]{Dongxin Yang}
\author[infn-pisa]{Saverio Citraro}
\author[infn-pisa]{Hikmat Nasimi}
\author[nut]{Jiandong Yu}
\author[nnvt]{Ge Jin}
\author[thu-dep]{Ming Zeng}
\author[nut]{Peng An}
\author[infn-pisa]{Luca Baldini}
\author[infn-pisa]{Ronaldo Bellazzini}
\author[infn-pisa]{Alessandro Brez}
\author[infn-torino]{Luca Latronico}
\author[infn-pisa]{Carmelo Sgr\`{o}}
\author[infn-pisa]{Gloria Spandre}
\author[infn-pisa]{Michele Pinchera}
\author[iaps]{Fabio Muleri}
\author[iaps]{Paolo Soffitta}
\author[iaps]{Enrico Costa}

\address[thu-dep]{Department of Engineering Physics, Tsinghua University, Beijing 100084, China}
\address[thu-doa]{Department of Astronomy, Tsinghua University, Beijing 100084, China}
\address[ihep]{Key Laboratory for Particle Astrophysics, Institute of High Energy Physics, Chinese Academy of Sciences, Beijing 100049, China}
\address[infn-pisa]{INFN-Pisa, Largo B. Pontecorvo 3, 56127 Pisa, Italy}
\address[infn-torino]{INFN, Sezione di Torino, Via Pietro Giuria 1, I-10125 Torino, Italy}
\address[nut]{School of Electronic and Information Engineering,  Ningbo University of Technology, Ningbo, Zhejiang 315211, China}
\address[nnvt]{North Night Vision Technology Co., Ltd., Nanjing 211106, China}
\address[iaps]{IAPS/INAF, Via Fosso del Cavaliere 100, 00133 Rome, Italy}

\begin{abstract}

\textit{PolarLight} is a compact soft X-ray polarimeter onboard a CubeSat, which was launched into a low-Earth orbit on October 29, 2018.  In March 2019, \textit{PolarLight} started full operation, and since then, regular observations with the Crab nebula, Sco X-1, and background regions have been conducted. Here we report the operation, calibration, and performance of \textit{PolarLight} in the orbit.  Based on these, we discuss how one can run a low-cost, shared CubeSat for space astronomy, and how CubeSats can play a role in modern space astronomy for technical demonstration, science observations, and student training. 

\end{abstract}

\begin{keyword}
\textit{PolarLight} \sep gas pixel detector \sep astronomy \sep X-ray \sep polarimetry \sep CubeSat
\end{keyword}

\end{frontmatter}


\section{Introduction}

Astronomical soft X-ray polarimetry in the energy range of a few keV is expected to break new grounds in understanding physics in the extreme universe \citep{Kallman2004,Soffitta2013}. Future missions with unprecedented sensitivity in this window have been planned, with the Imaging X-ray Polarimetry Explorer (IXPE)  scheduled to launch in 2021 \citep{Weisskopf2016}, and the enhanced X-ray Timing and Polarimetry (eXTP) several years later \citep{Zhang2019}.  Both missions will carry high-sensitivity soft X-ray polarimeters based on the gas pixel detector \citep[GPD;][]{Costa2001,Bellazzini2013}. To have a direct flight test of the new technique, a detector onboard a CubeSat was launched into a Sun-synchronous orbit on 29 October 2018, named \textit{PolarLight} \citep{Feng2019}.  With observations of  the Crab nebula in the first year, \textit{PolarLight} has re-detected X-ray polarization consistent with that obtained more than 40 years ago with the OSO-8 experiment \citep{Weisskopf1976,Weisskopf1978a}, and also discovered a time variation in polarization at a 3$\sigma$ level associated with the pulsar emission \citep{Feng2020a}.

\textit{PolarLight} is the name of a payload, not a spacecraft. It is not even the sole payload on the spacecraft, while others have no specific requirements for attitude control.  The key component of \textit{PolarLight} is a GPD mounted on a printed circuit board, assisted with a high voltage (HV) board and a data acquisition board. The three boards stack vertically and occupy a standard unit of the CubeSat.  The field of view (FOV) of \textit{PolarLight} is shaped by a capillary collimator with round holes to have a full width at half maximum (FWHM) of 2.3$^\circ$.  The sensitive area of the detector has a size of 1.4~cm $\times$ 1.4~cm. Incident X-rays are attenuated at energies below 2~keV by a thin layer of thermal coat surrounding the satellite and the beryllium window of the detector. GPD is a 2D position sensitive gaseous proportional counter. Following the absorption of an X-ray in the sensitive volume, a 2D image of the photoelectron trajectory projected on the detector plane is obtained.  A statistical analysis of the initial direction of photoelectrons derived from the image tells the polarization of X-rays. The pixels have a hexagonal pattern with a pitch of 50~$\mu$m. The arrival time of each event is tagged by an internal timer synchronous with the global positioning system (GPS). Electron drift in the chamber limits the absolute timing accuracy to be about 1 $\mu$s. The effective energy range for polarimetry is 2--8 keV, with an energy resolution of about 16\% (${\rm FWHM}/E$) at 6 keV, typical of gas proportional counters.  More details of the instrument and its ground calibrations can be found in \citet{Feng2019}. 

In this paper, we describe the operation, calibration, and performance of \textit{PolarLight} in the orbit. As there is increasing interest in applications of CubeSats for space astronomy, we will discuss how to efficiently run a commercial CubeSat for high energy astronomy, and specifically, how CubeSats can play a role in modern space astronomy for technical demonstration, science observations, and student training.

\section{Operation}

\textit{PolarLight} resides in a 6U CubeSat called Tongchuan-1, manufactured by the company Spacety Co.\ Ltd\footnote{\url{http://www.spacety.com}}.  The CubeSat was launched from the Jiuquan Satellite Launch Center into a nearly circular Sun-synchronous orbit with an altitude of 520~km and a period of about 95~minutes. The CubeSat manufacturer is also responsible for telemetry and data transmission of the spacecraft. We provide commands needed for the payload and associated satellite control. 

\subsection{Observations}

\begin{figure}[t]
\centering
\includegraphics[width=\columnwidth]{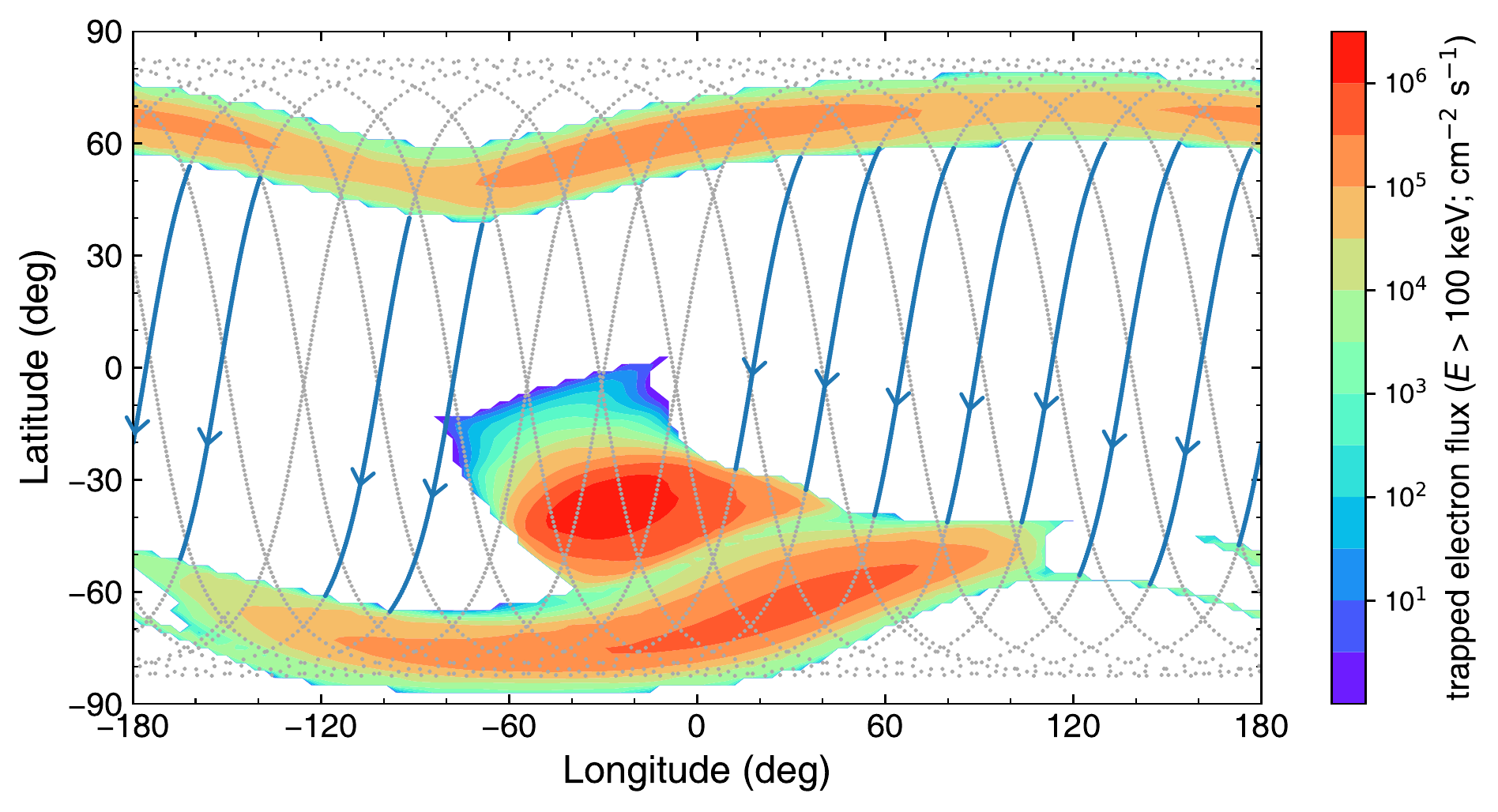}
\caption{A typical one-day (July 30, 2019) orbit of \textit{PolarLight}. The colormap shows the flux of  trapped electrons with energies above 100~keV in the orbit. The blue lines with arrows indicate orbital trajectories during which \textit{PolarLight} is observing the Crab nebula.
\label{fig:orbit}}
\end{figure}

\begin{figure}[h!]
\centering
\includegraphics[width=\columnwidth]{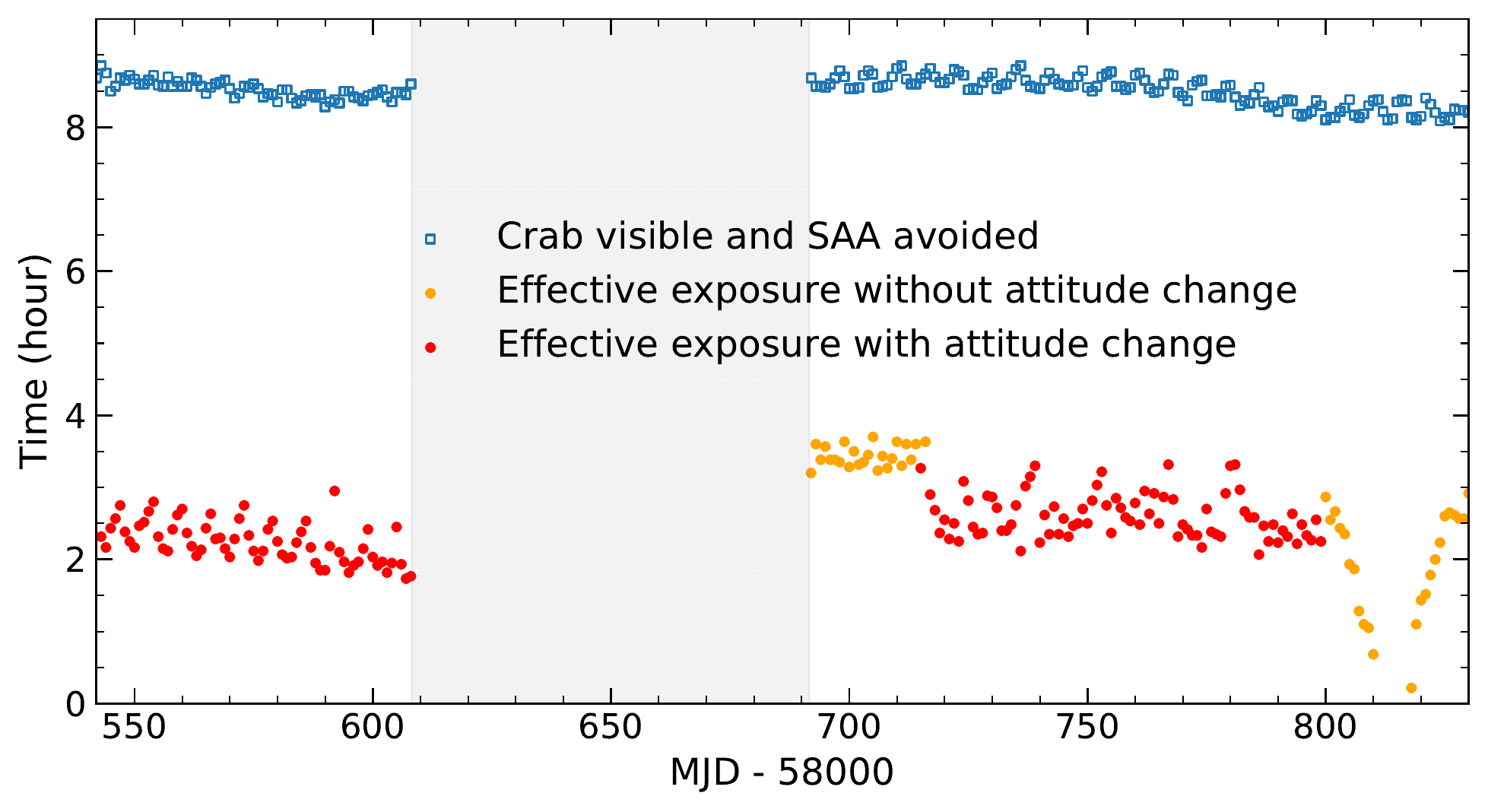}
\caption{Maximum possible (blue squares) and actual (yellow and red dots) observing time per day over a year for the Crab nebula. The blue squares indicate the total time in a day that the Crab nebula is visible to the CubeSat and the CubeSat is outside of the regions with trapped charged particles (SAA and the two polar regions, see Figure~\ref{fig:orbit}).  The dots indicate the actual exposures in which science observations are executed, with observable windows less than 15 minutes discarded, and the HV ramp up/down time excluded. The difference between yellow and red dots is that a fixed attitude is used in the former while an attitude change is applied in the latter, which may cost several minutes.  In MJD 58800--58830, a drop of actual exposure times is due to the fact that many observable windows are shorter than than 15 minutes and thus discarded.  The gray region indicates the time interval where no observations are scheduled due to Sun avoidance. 
\label{fig:exposure}}
\end{figure}

In the first four months after launch,  the CubeSat underwent a commissioning phase with major upgrades of software, and the \textit{PolarLight} detector was briefly powered on for functional test. Regular science observations with \textit{PolarLight} started in March 2019. Figure~\ref{fig:orbit} shows a typical one-day orbit of the CubeSat on top of the flux map of high energy ($>100$~keV) electrons trapped by the Earth's magnetosphere in the orbital plane. The map is generated from SPENVIS\footnote{\url{http://www.spenvis.oma.be}}. As charged particles may cause damage to the GPD if the flux is too high, the HV power supply is powered off when the CubeSat passes the high flux regions, including the south Atlantic anomaly (SAA) and two polar regions (Figure~\ref{fig:orbit}).  The time needed from the spacecraft leaving the high flux region to data acquisition is 230~s, including the HV ramping up and stabilizing time, and some margin in case the boundary of the high flux region is varying. Similarly, the data acquisition stops a similar amount of time prior to entering the high flux region.  As a consequence, we do not attempt to power on the detector if the window for observation is not long enough.  In practice, with such a constraint, we only use observing windows that are longer than 15 minutes and discard those near the two poles.  In addition, in around half of the durations suitable for observations, the target may be occulted by the Earth. On average, there are around 10 orbits a day, each with an effective exposure of 15 minutes or so, in which science data can be obtained.  The effective exposure times per day for the Crab nebula is plotted in Figure~\ref{fig:exposure}.

The Crab nebula, the only source with a significant detection in soft X-ray polarization \citep{Weisskopf1976,Weisskopf1978a}, is the primary target of \textit{PolarLight}.  From May 11 to July 10, 2019, the Crab was too close to the Sun on the sky plane to schedule any observations. During that period, the brightest persistent X-ray source Sco X-1 was selected as the target. 

During Earth occultation of the target, the in-orbit background is measured, by pointing at sky regions without bright sources or the Earth's atmosphere.  As the background is dominated by charged particles in the orbit, the pointing is not crucial. 

The operation of the satellite and detector such as attitude control, data acquisition, storage and transfer, and HV regulation are done via commands stored in the buffer of the onboard computer (OBC).  The buffer can accept 128 pieces of commands in total. We generate and upload the commands to the CubeSat everyday for the operation in the next day according to the orbit predicted using the two-line element sets\footnote{obtained from \url{https://www.space-track.org}}.  

\begin{figure}[t]
\centering
\includegraphics[width=\columnwidth]{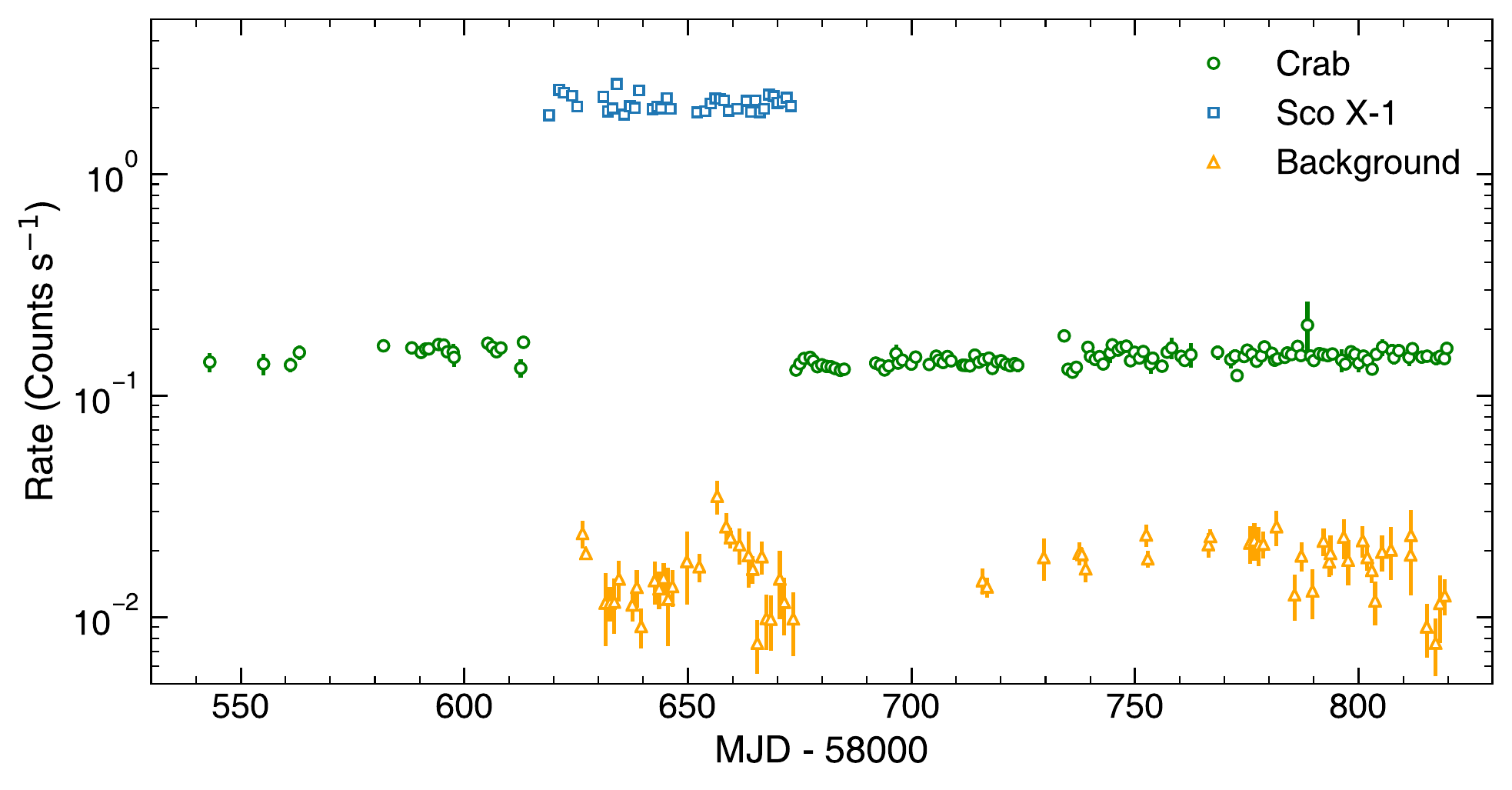}
\caption{One-day averaged lightcurves of the Crab, Sco X-1, and background observed with \textit{PolarLight} in the energy range of 2--8 keV.
\label{fig:lc}}
\end{figure}

We note that the efficiency of observation varies with time, mainly due to occasional issues with the CubeSat, e.g., problems associated with the OBC, global positioning system (GPS), or star tracker.  Sometimes, the application softwares of the OBC or payload computer needed to upgrade, which took several days or even weeks. The one-day averaged lightcurves of the Crab, Sco X-1, and background are shown in Figure~\ref{fig:lc}.

\subsection{Attitude control}

The satellite can be stabilized in two modes, controlled by the reaction wheels or magnetic torquer.  Scientific observations utilize the wheel control mode, in which case the star tracker is needed for precise attitude measurement. There is only one star tracker mounted on the CubeSat. \textit{PolarLight} points at the $Y$ direction and the star tracker points at $Z$.  The star tracker must avoid the sun by at least 60$^\circ$ and the Earth by at least 92$^\circ$ in order to work. This largely constrains the orientation of the CubeSat given a target. In practice, it often occurs that an observation cannot be finished with a fixed attitude during an observing window. In that case, we have to roll the satellite at some point during an observation, which leads to a waste of effective exposure for a few minutes until the pointing is stabilized again (see red dots in Figure~\ref{fig:exposure}). 

As mentioned above, we observe the background when the target is occulted by the Earth. In the meanwhile, to charge the battery and also extend the lifetime of reaction wheels, the satellite is controlled in the magnetic mode, in which the CubeSat solar panel points at the Sun but rocks with a half angle of about 30$^\circ$. As a consequence, the star tracker's Sun/Earth constraint is violated, so an accurate attitude of the satellite cannot be obtained for background observations.  In practice, we do not need accurate knowledge of attitude for background observations, because the background is dominated by high energy charged particles which are nearly isotropic in the space. To evaluate the modulation caused by background, one would also use the detector coordinates instead of the sky coordinates.

\subsection{Telemetry and Data transfer}

The status of the CubeSat, including the attitude, timing, temperatures, voltages, etc., is acquired and recorded by the payload computer every ten seconds.  \textit{PolarLight} has a flash memory of 64~Mbytes on the data acquisition board. When observing the Crab, science and housekeeping data with a total amount of $\sim$30~Mbytes are produced everyday.  The data are first stored in the flash of the payload during observations, and then transferred to the payload computer via a serial peripheral interface, before the flash is erased at the end of the day.  For Sco X-1, the flash is not large enough to support a full day observation. Once the flash memory is full, the data are transferred to the payload computer followed by an erase. These two steps (transfer and erase) take about an hour for a full flash.  The payload computer has an eMMC memory of 32~Gbytes, enabling temporary data storage for a sufficiently long time even for observations with Sco X-1. The data are transmitted to the ground station every other day via the X band.  Each time a data package of about 100~Mbytes can be downloaded. If it is missed or unsuccessful,  data transmission could occur twice a day when the satellite is seen by the ground station with an elevation angle higher than 40$^\circ$.  Satellite telemetry is allowed  4--6 times a day via the UHF channel when the CubeSat flies over another 4 ground stations in different locations.

\section{In-flight calibrations and performance}

\subsection{Pointing calibration}

\begin{figure}[t]
\centering
\includegraphics[width=0.8\columnwidth]{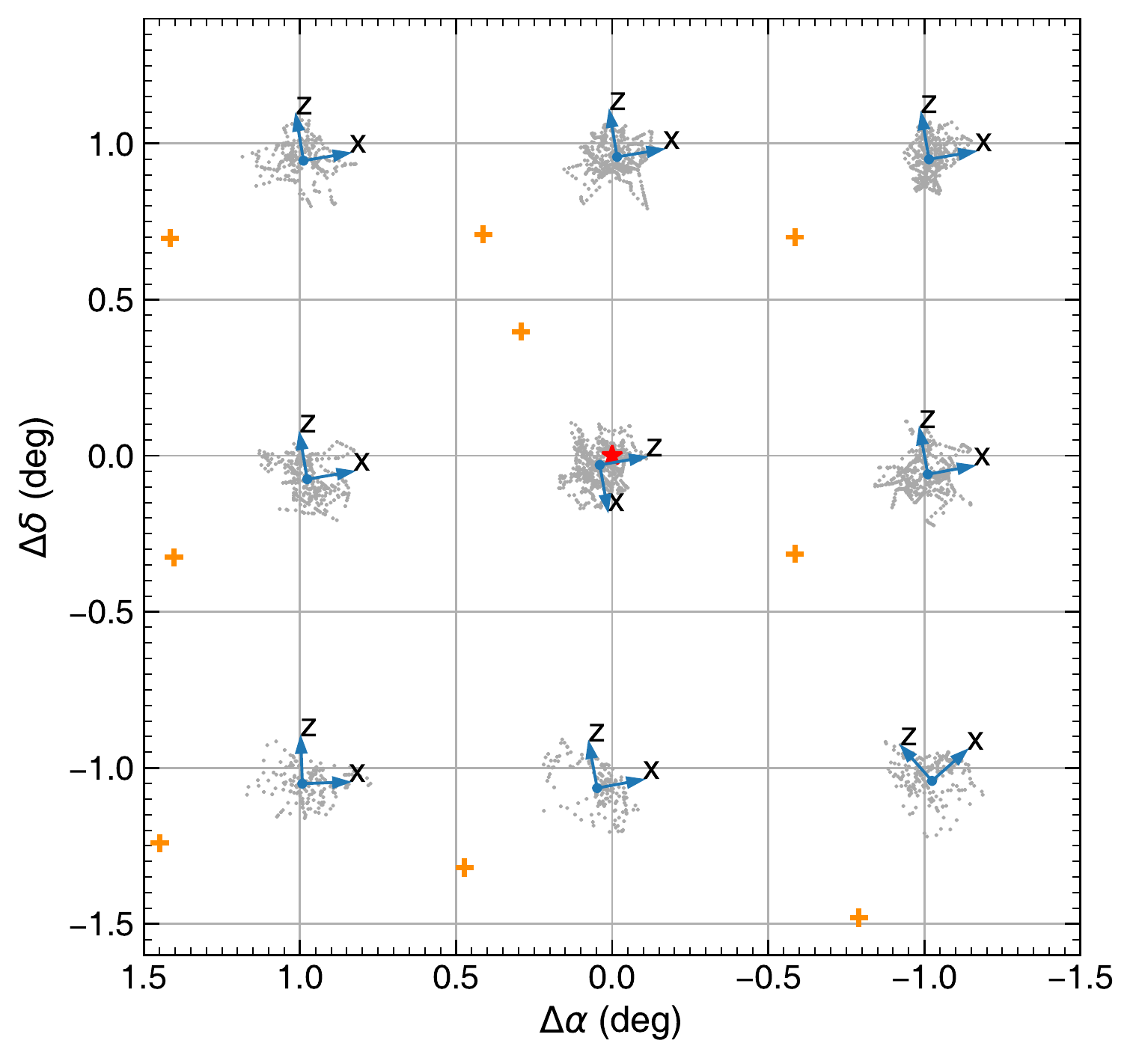}
\caption{Pointed observations surrounding Sco~X-1 for pointing calibration. The gray dots show the pointing direction ($Y$) measured with the star tracker. The blue axes mark the $X$ and $Z$ directions projected on the sky plane. The orange crosses indicate the best-fit optical axis of the collimator. The red star indicates the position of Sco~X-1.  
\label{fig:scan}}
\end{figure}

\begin{figure}[t]
\centering
\includegraphics[width=0.8\columnwidth]{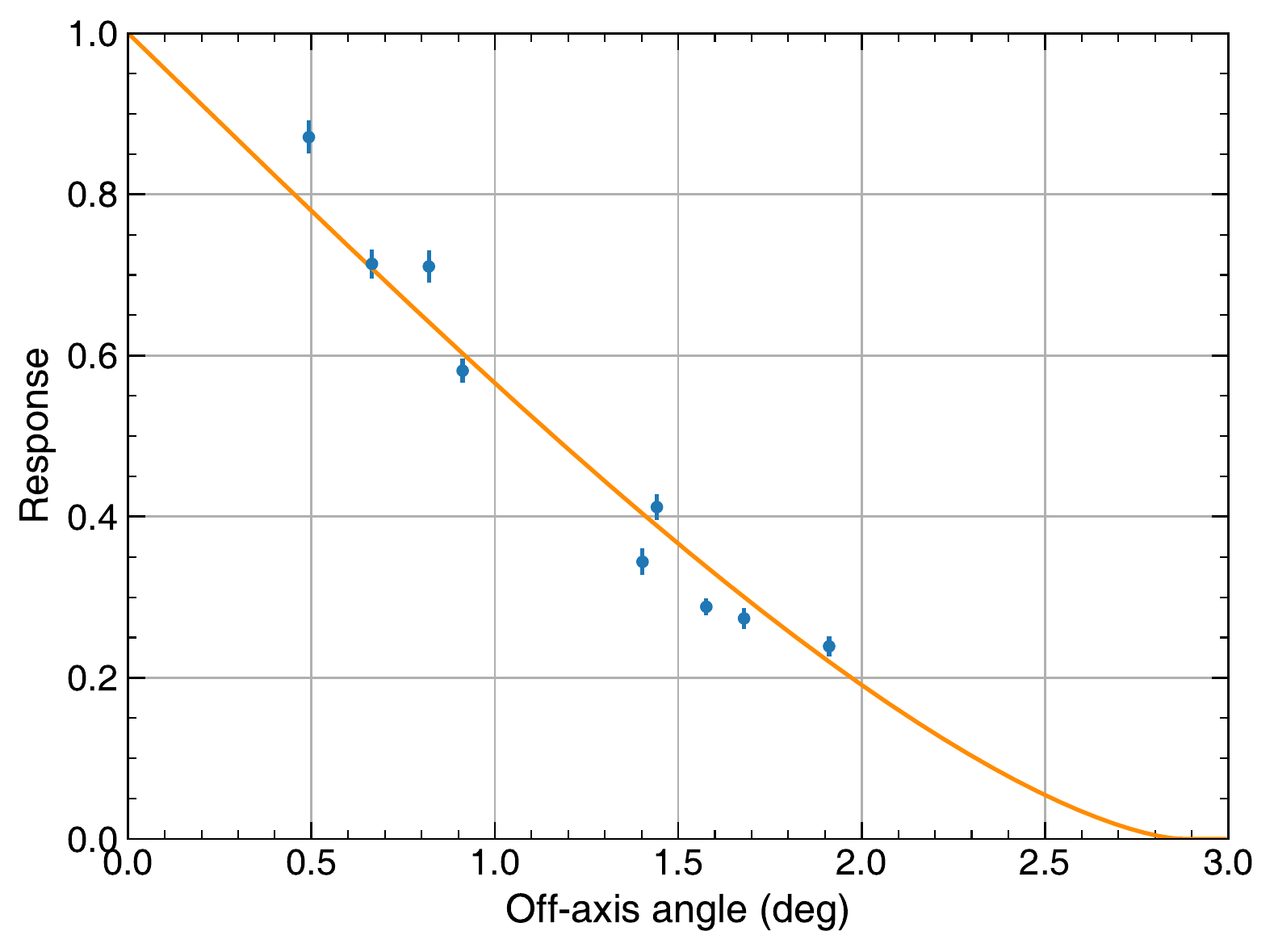}
\caption{Normalized Sco X-1 count rate as a function of the off-axis angle after the pointing calibration.  The line indicates the angular response of the collimator. 
\label{fig:scan_resp}}
\end{figure}

The collimator is mounted on the top surface of the detector and points to the $Y$ direction nominally. The deviation of the optical axis to the nominal direction in the reference frame determined by the star tracker needs be calibrated after launch.  Sco~X-1 is used for the calibration thanks to its brightness.  We performed nine pointed observations in a sky region surrounding Sco X-1, in a pattern of $3 \times 3$ with a spacing of 1$^\circ$ (Figure \ref{fig:scan}).  The flux from Sco X-1 is time variable, and is corrected with the flux measured with MAXI \citep{Matsuoka2009}. The collimator has round holes and its angular response on flux is axially symmetric and well understood \citep{Feng2019}.  Given the measured fluxes and the angular response of the collimator, the pointing of the collimator or its misalignment with respect to the $Y$ direction can be inferred.  The best-fit result indicates that the collimator's optical axis is misaligned with respect to the $Y$ axis by $0.50^\circ \pm 0.07^\circ$, with a roll angle of $111^\circ \pm 8^\circ$ counter-clockwise from the $Z$ direction. Correcting the misalignment will find back 20\% of the source flux. The measured fluxes from Sco X-1 as a function of the off-axis angle with respect to the best-fit pointing is shown in Figure~\ref{fig:scan_resp}.

\begin{figure}[t]
\centering 
\includegraphics[width=0.8\columnwidth]{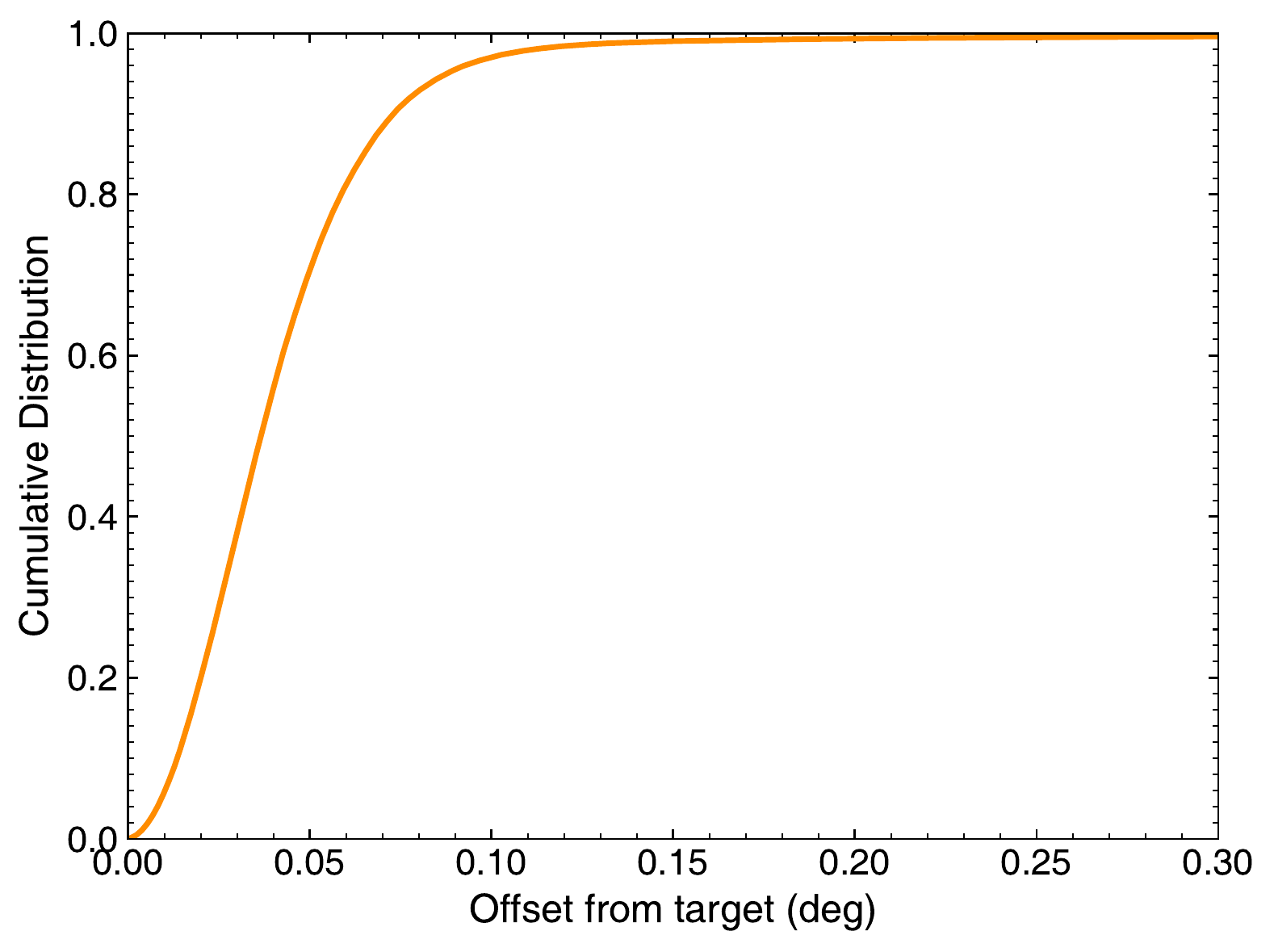}
\caption{Cumulative distribution of the off-axis angle during pointed observations. The distribution is constructed from single attitude measurements sampled at a frequency of 0.1~Hz over all observations. The pointing is within 0.07$^\circ$ around the target over 90\% of the observing time, or or 0.15$^\circ$ over 99\% of time.
\label{fig:pointing}}
\end{figure}

When \textit{PolarLight} is observing a target, the cumulative distribution of the off-axis angle is shown in Figure~\ref{fig:pointing}.  This reflects the stability of attitude over time. As one can see, in 99\% of time the pointing is within 0.15$^\circ$ of the target. In data analysis, we recommend to use a threshold of 0.2$^\circ$ to select on-axis data. This encircles 99.3\% of data and such a small angle has no effect to the polarization analysis. 

\subsection{Energy calibration}

There is no radioactive source for calibration on \textit{PolarLight}. The measured energy spectrum does not show any line feature.  The detector window and thermal coat absorb low energy photons, resulting in a peak around 2 keV. This feature, along with the fact that the Crab spectral shape in this energy band is nearly constant,  can be used to find the relation between the pulse height amplitude (PHA) in unit of analog-to-digital converter (ADC) number and photon energy.  The Crab spectrum is adopted from \citet{Kirsch2005} and used to simulate a measured energy spectrum using Geant4. The energy resolution measured in the lab \citep{Feng2019} is taken into account, and is found insensitive to the gain calibration.  For Sco X-1, we adopt the energy spectrum in the non-flare state \citep{Church2012} as the input. We note the caveat that the Sco X-1 spectral shape may vary so that the energy calibration should be used with caution. 

\begin{figure}[t]
\centering
\includegraphics[width=0.8\columnwidth]{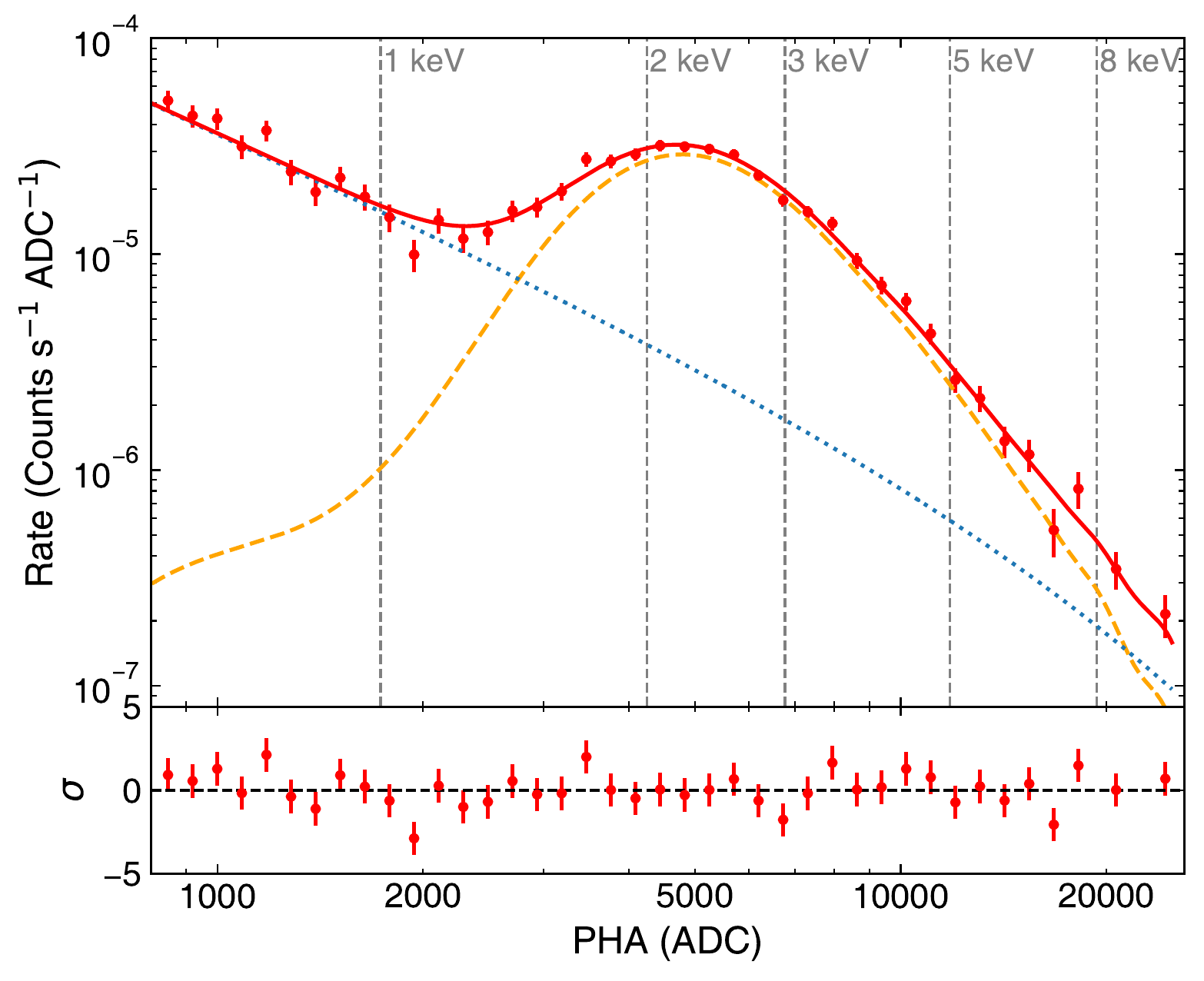}
\caption{PHA spectrum of the Crab observed with \textit{PolarLight}. The orange dash line indicates the source model simulated using Geant4. The blue dotted line indicates the background component described by a cutoff power-law model, with the power-law index ($b_1$) and cutoff PHA ($b_2$) derived from the pure background spectrum. 
\label{fig:fitspec}}
\end{figure}

The measured PHA spectrum constructed from ``X-ray events'' selected using a simple algorithm\footnote{The ``X-ray events'' are defined as those whose track image has a diagonal of no more than 70 pixels, an eccentricity not higher than 50, and only one isolated charge island.} \citep{Feng2020a} is shown in Figure~\ref{fig:fitspec}. It consists of two components, respectively, from the source and background.  The background component shows an exponentially cutoff power-law shape, $b_0 \times {\rm PHA}^{b_1} e^{-{\rm PHA} / b_2}$, revealed by observations of the background region. The source component is translated from the simulated energy spectrum by relating the PHA with energy, $E = a_0 + a_1 \times {\rm PHA}$. Here $a_{0,1}$ and $b_{0,1,2}$ are coefficients to be determined. We first fit the pure background spectrum to find $b_{0,1,2}$, and then fix $b_{1,2}$ when fitting the Crab spectrum, leaving $b_0$ and $a_{0,1}$ as free parameters.   The coefficient $b_0$ is linked to the total background flux, which is displayed in Figure~\ref{fig:lc}. 

\begin{figure}[t]
\centering
\includegraphics[width=0.8\columnwidth]{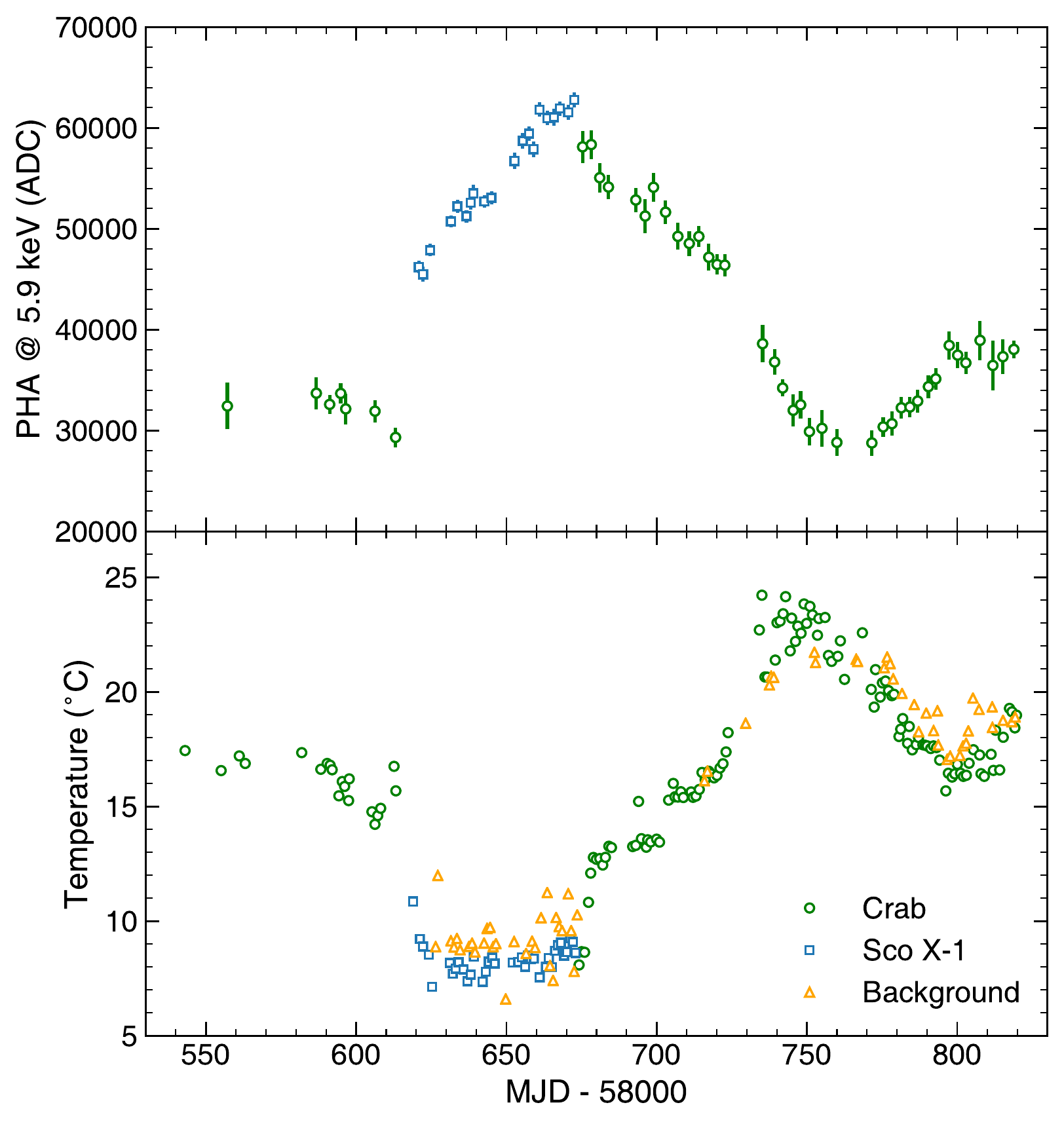}
\caption{Gain (top) and temperature (bottom) of \textit{PolarLight} in the orbit as a function of time.  The gain is found by fitting the measured PHA spectrum with a simulated energy spectrum, and is indicated here by the peak position in the PHA spectrum resulted from 5.9~keV X-rays at a HV of 3200~V. The temperature is mainly determined by the orientation of the CubeSat with respect to the Sun. 
\label{fig:gain_temp}}
\end{figure}

The data in every 2--4 days (with a similar number of photons) are grouped to investigate the time variation of the detector gain, which is calculated using the above recipe and shown in Figure~\ref{fig:gain_temp}. The detector gain is indicated as the peak position in the PHA spectrum for 5.9~keV X-rays ($^{55}$Fe K$\alpha$), using the inferred PHA-energy relation.  In the orbit, the detector is operated at a HV of 3000~V.  Here in Figure~\ref{fig:lc}, we have multiplied a factor of 3.4 to the PHA as if it was measured at a HV of 3200~V to enable direct comparisons with the gain measurements in the lab before the launch \citep{Feng2019}.

There is no active thermal control for the spacecraft. The temperature of \textit{PolarLight} measured from a sensor on the data acquisition board in the space as a function of time is shown in Figure \ref{fig:gain_temp}. The temperature, varying in a range of 11$^\circ$C to 24$^\circ$C during observations of the Crab nebula,  is mainly determined by the solar angle. The gain is found to be anti-correlated with the temperature, with a time delay of roughly 10--20 days.  This is also observed in the lab, but its nature is still uncertain and under investigation. 

\begin{figure}[t]
\centering
\includegraphics[width=0.8\columnwidth]{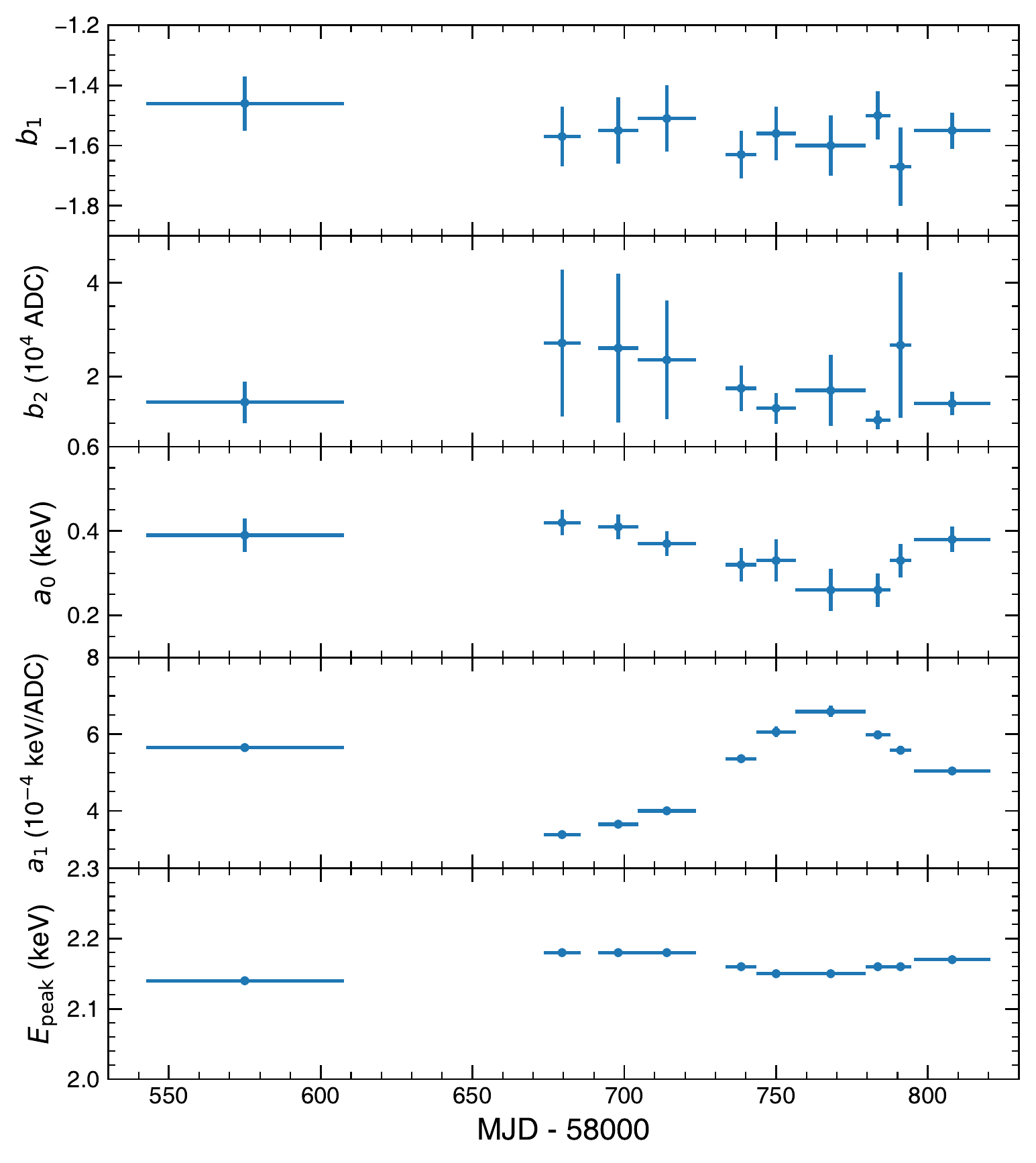}
\caption{Spectral parameters obtained from energy calibrations with the Crab nebula in different epochs. Parameters $b_{1,2}$ are to describe the background spectral shape, and $a_{0,1}$ are linear coefficients linking PHA and energy. $E_{\rm peak}$ is the peak location around 2~keV in the energy spectrum of the Crab nebula (Figure~\ref{fig:fitspec}). The coefficient $b_0$ that indicates the background level is shown in Figure~\ref{fig:lc}.
\label{fig:variability}}
\end{figure}

Depending on the gain variation shown in Figure~\ref{fig:gain_temp}, we then group the observations into different epochs for energy calibration, with each epoch containing observations with a nearly constant gain. The spectral parameters obtained from energy calibrations are displayed in Figure~\ref{fig:variability}. The coefficient $a_1$ indicates the detector gain, showing a trend consistent with that displayed in Figure~\ref{fig:gain_temp}. The coefficient $a_0$ is the intercept energy and is anti-correlated with the gain. This is likely because different noise cuts are used at different gains. The combination of the two parameters, however, produces a nearly constant peak location ($E_{\rm peak}$) in the energy spectrum, the local maximum just above 2 keV due to window absorption of the source spectrum. The peak-to-peak variation of $E_{\rm peak}$ is less than 2\% and the standard deviation is 0.6\%, justifying the accuracy of calibration.  The parameters $b_{1,2}$ indicate the background spectral shape, and are constant over time within errors.

\subsection{Background and particle discrimination}

\begin{figure*}
\centering
\includegraphics[width=0.24\textwidth]{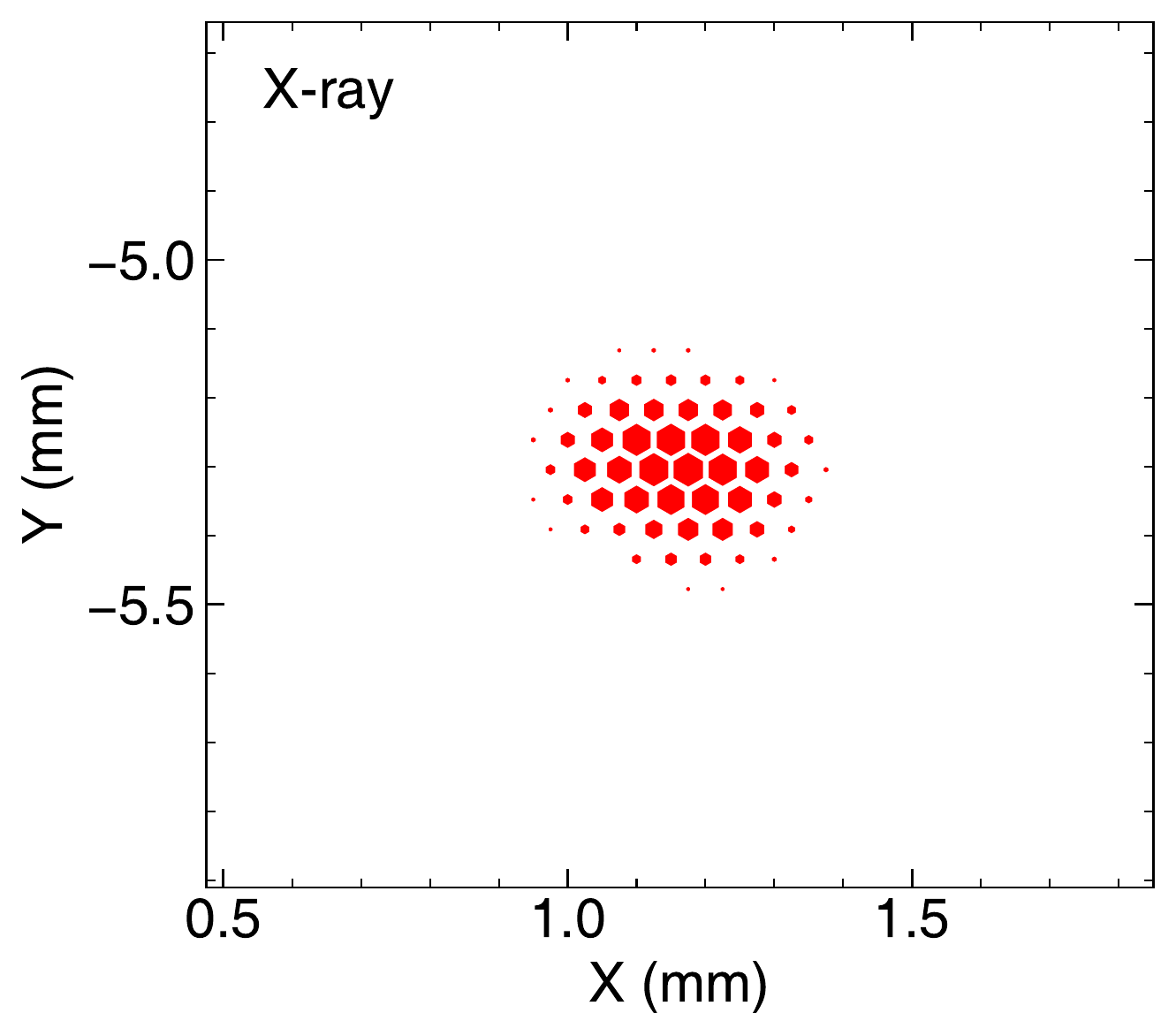}
\includegraphics[width=0.23\textwidth]{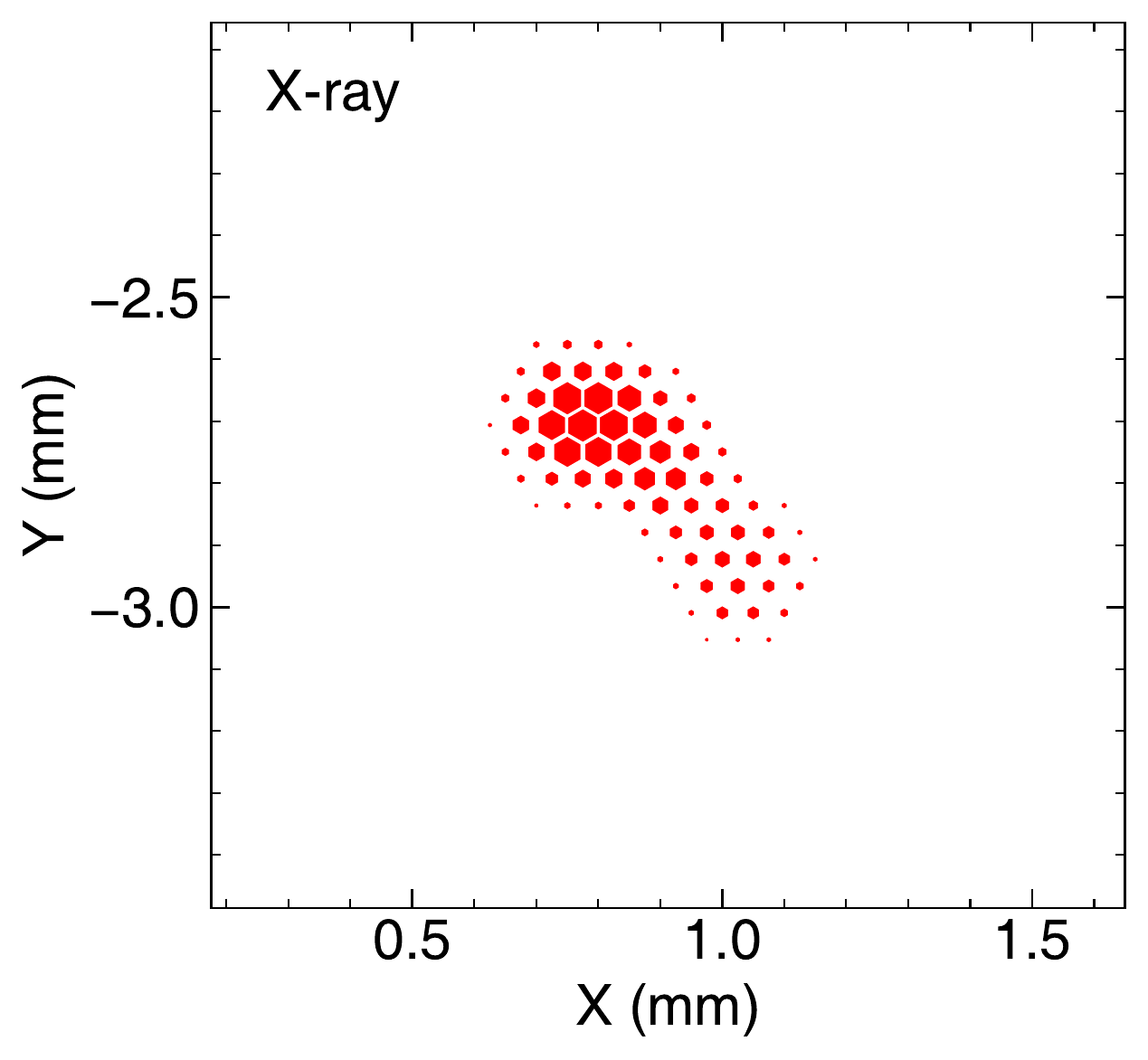}
\includegraphics[width=0.27\textwidth]{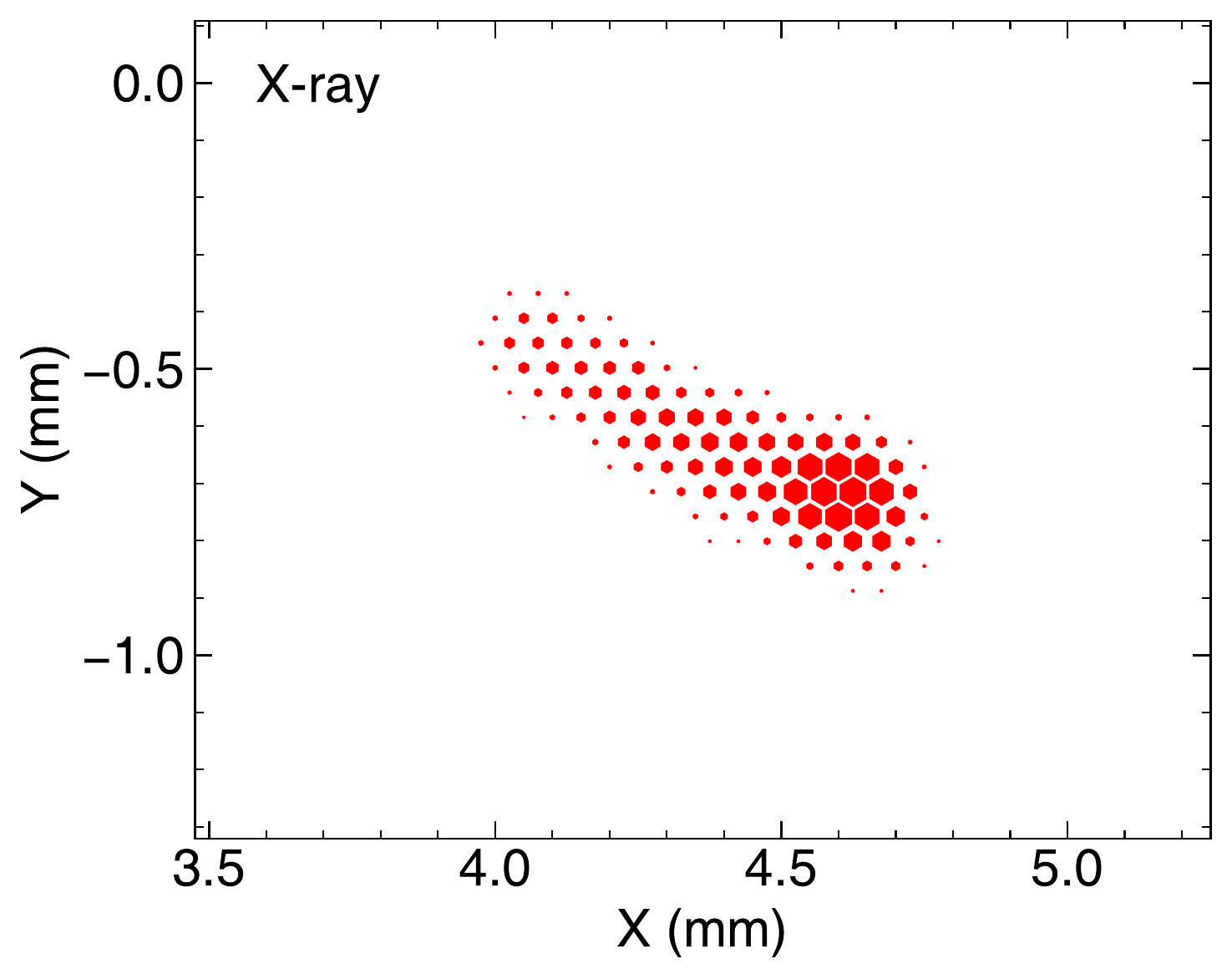}
\includegraphics[width=0.21\textwidth]{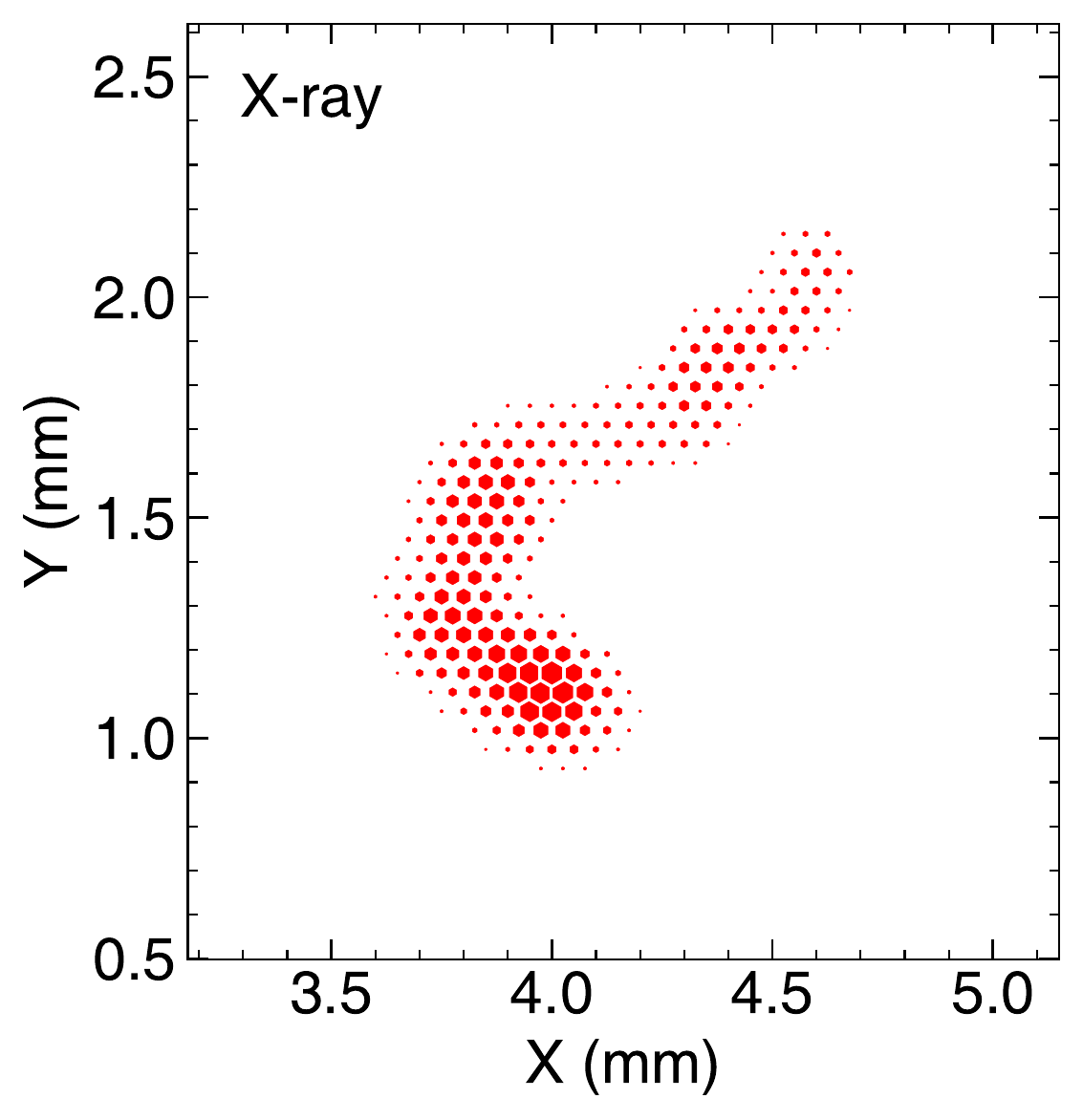}\\
\includegraphics[width=0.43\textwidth]{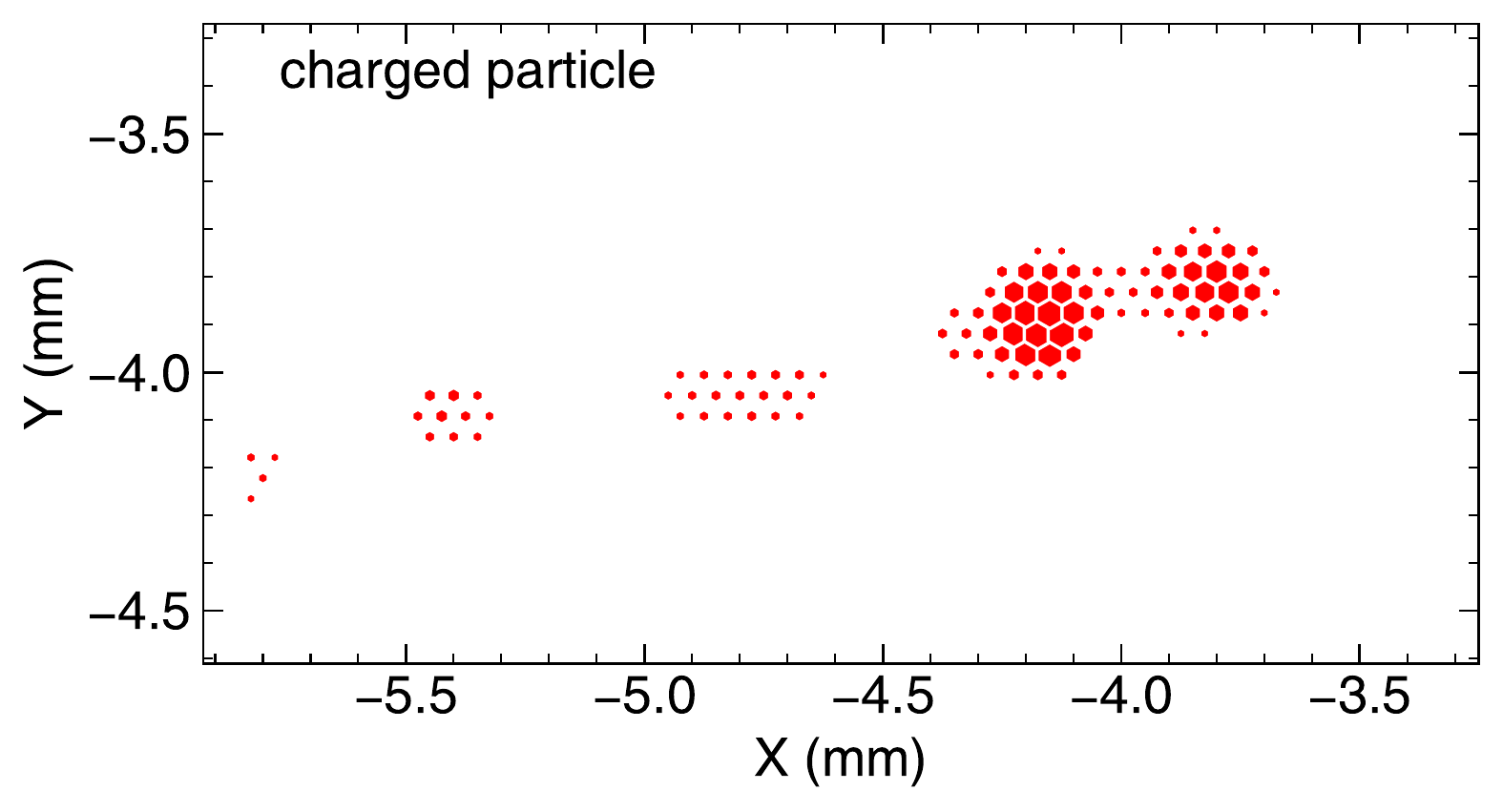}
\includegraphics[width=0.23\textwidth]{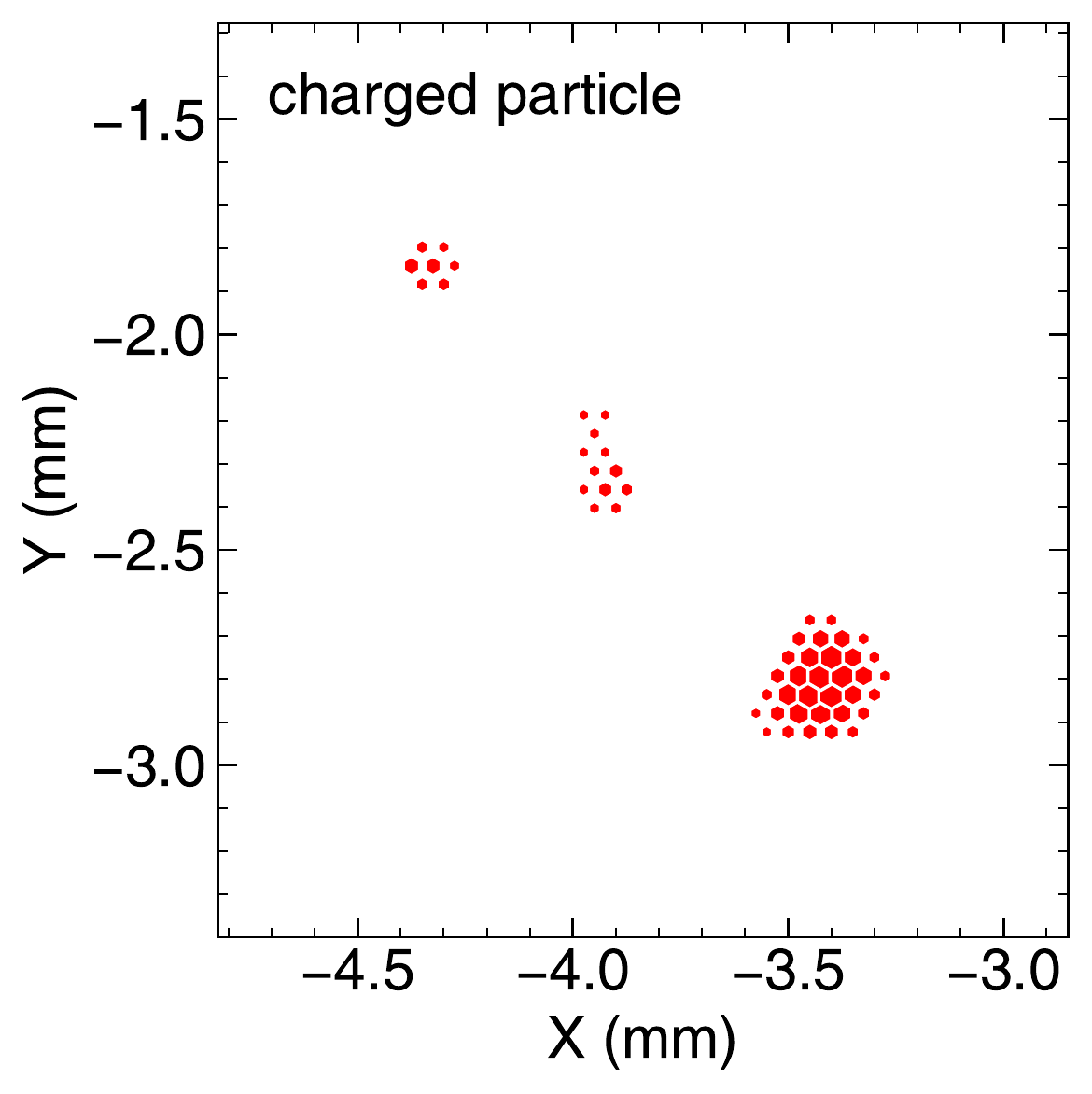}
\includegraphics[width=0.26\textwidth]{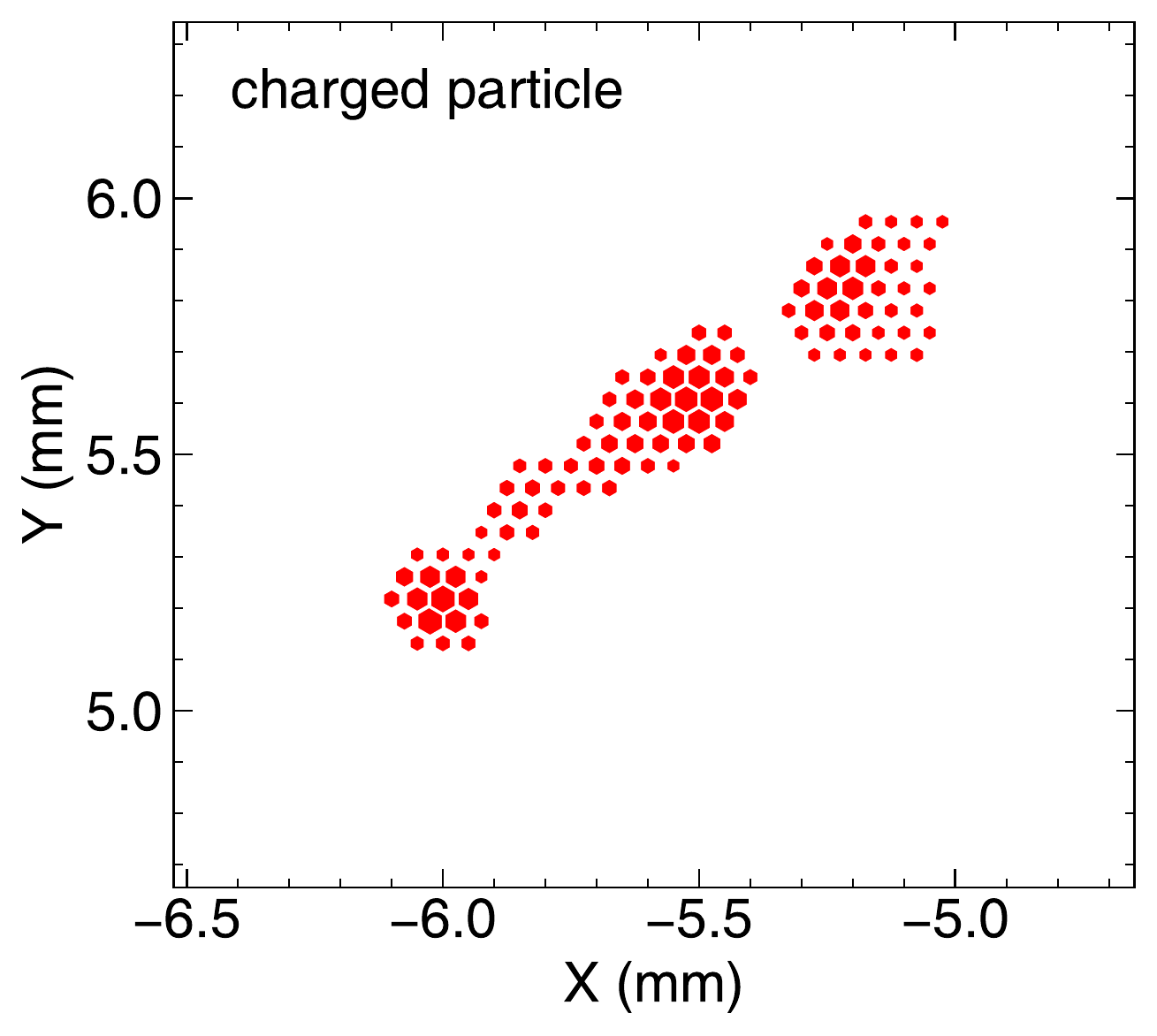}
\caption{Typical track images produced by X-ray events (top) or charged particle events (bottom).  The images reflect the energy loss of charged particles in the GPD projected on the detector plane with hexagonal pixels at a pitch of 50 $\mu$m. Electrons of a few keV usually leave behind a curved track, while high energy electrons may produce long, straight tracks. 
\label{fig:track}}
\end{figure*}

Events due to high energy charged particles and X-rays in the energy band of our interest could be distinguished based on their different track morphologies. In general, high energy charged particles tend to result in long, straight tracks, while the X-rays or electrons in the energy band of our interest produce relatively short and curved tracks.  However, high energy charged particle may also yield secondary electrons or X-rays in the energy band of our interest, and produce a fraction of events that are indistinguishable. Figure~\ref{fig:track} shows some typical track images produced by X-rays and charged particle events; or precisely, some images that are most likely produced by X-rays or charged particles, based on knowledges gained from laboratory tests and Geant4 simulations.  Owing to the narrow FOV, the background of \textit{PolarLight} is dominated by charged particles in the orbit, rather than the cosmic X-ray background or albedo X-rays from the Earth's atmosphere.  A detailed modeling and analysis of the background will be reported elsewhere (Huang et al.\ 2020, in prep).  

\section{Discussion}

At the time of writing, \textit{PolarLight} has been working in space for more than a year. The detector is always in good health and there is no observable performance degradation.  The purpose of the project is to have a direct demonstration of the new technique for soft X-ray polarimetry in space, and has been fulfilled. In addition, observations of the Crab nebula has produced interesting science results \citep{Feng2020a}. A direct measurement of the in-orbit background can bring a better understanding of the instrument sensitivity and systematics, which will be useful for future missions like IXPE and eXTP. 

In recent years, along with a rapid growth of commercial aerospace, applications of CubeSats in space astronomy \citep{Shkolnik2018} have become more and more frequent.  Unlike many other astronomical CubeSats, the case for \textit{PolarLight} is somewhat special. Many astronomical CubeSats have a dedicated spacecraft, while \textit{PolarLight} shares the satellite with other payloads. This, however, has almost no effect on its operation, because the operation of other payloads on the same CubeSat does not involve attitude control in most cases.  On the other hand, a shared CubeSat lowers the cost substantially.  For a dedicated satellite, routine operations can be programed into the OBC to maximize the efficiency.  The CubeSat for \textit{PolarLight} is a standard product of the company, without any customized changes to the hardware or software infrastructure.  This means that we have to create all the operation commands at a lower level and do this every day.  However, we find that such a deep involvement into  CubeSat operation has become a good way for student training, see discussions below.  In the CubeSat, off-the-shelf components are used and can satisfy the requirements for \textit{PolarLight}.  For high energy astrophysical payloads similar to \textit{PolarLight}, there should be a wide range of CubeSat products available in the market.

Taking \textit{PolarLight} as an example, we outline three aspects of advantages about the application of CubeSats in modern space astronomy --- how nano-satellites can play a role nowadays especially when astronomical observatories have become larger and larger. 

The most obvious application of CubeSats is for technical demonstration. Compared with sounding rockets and balloons, CubeSats are much smaller but can fly longer, with a lifetime of several years.  The price of a CubeSat keeps decreasing in recent years, thanks to the maturity of industry and mass production. Therefore, CubeSats have become a favorable platform to test small-scale instruments. For \textit{PolarLight}, as mentioned above, a shared CubeSat minimizes the expense. It makes a space experiment affordable by a small research group, and no support is needed from the space agency.  This can essentially speed up technical innovations and iterations in future space experiments.

Modern astronomy requires huge telescopes with unprecedented sensitivity. It is impossible to implement these kinds of instruments on CubeSats. Major observatories should benefit a large community. Their observing times are usually shared by many astronomers. It becomes unpractical for large observatories to monitor a particular source for a very long time. Thus, such dedicated, highly customized observation programs that cannot be accomplished by large facilities may be supplemented with CubeSats.  For example, \textit{PolarLight} has been monitoring the Crab nebula for more than a year, and will continue to do that in the rest of its lifetime. These results, besides valuable and interesting on their own right, can also help optimize observation programs or trigger programs of opportunity for future larger missions like IXPE or eXTP.

Last but not least, CubeSats can be used as a great tool for student training. The development of large-scale missions may take a decade or so.  It is impossible for a student to follow the project from the beginning to end. It is also hard for a student to take a leading role in those projects.  These, however, can be well adapted into a CubeSat project.  For the example of \textit{PolarLight}, it took one year for instrument development (from a laboratory setup to a calibrated payload),  one to three years for in-orbit operation, and in the meanwhile, one year or two for data analysis and result publications. These make up a well-defined thesis project for a PhD student.  The daily operation of \textit{PolarLight} described above is conducted by a PhD student. With such a thesis topic, the student is trained in both science and engineering, as well as leadership. 

\section*{Acknowledgements}

We thank the anonymous referees for their useful comments. HF acknowledges funding support from the National Natural Science Foundation of China under the grant Nos.\ 11633003 \& 11821303, the CAS Strategic Priority Program on Space Science (grant No.\ XDA15020501-02), and the National Key R\&D Project (grants Nos.\ 2016YFA040080X \& 2018YFA0404502).  


\end{document}